# Would a File by Any Other Name Seem as Malicious?

Andre T. Nguyen
Booz Allen Hamilton
UMBC
Nguyen_Andre@bah.com

Edward Raff
Booz Allen Hamilton
UMBC
Raff_Edward@bah.com

Aaron Sant-Miller
Booz Allen Hamilton
Sant-Miller_Aaron@bah.com

## ABSTRACT

Successful malware attacks on information technology systems can cause millions of dollars in damage, the exposure of sensitive and private information, and the irreversible destruction of data. Anti-virus systems that analyze a file's contents use a combination of static and dynamic analysis to detect and remove/remediate such malware. However, examining a file's entire contents is not always possible in practice, as the volume and velocity of incoming data may be too high, or access to the underlying file contents may be restricted or unavailable. If it were possible to obtain estimates of a file's relative likelihood of being malicious without looking at the file contents, we could better prioritize file processing order and aid analysts in situations where a file is unavailable.

In this work, we demonstrate that file names can contain information predictive of the presence of malware in a file. In particular, we show the effectiveness of a character-level convolutional neural network at predicting malware status using file names on Endgame's EMBER malware detection benchmark dataset.

## 1 INTRODUCTION

Organizations in government and industry increasingly rely on information technology (IT), digitized data, and networked computer assets. Consequently, they face a growing risk of exposure to cyber attacks (i.e., malicious attempts at stealing, altering, or destroying IT systems and data). Yet, as computer networks grow in size, so do the challenges cybersecurity professionals face in securing them. With more connected devices, more users, and more complex systems, adversarial attack opportunities increase exponentially. In 2018, on average, large organizations took almost 200 days to identify a cyber breach and almost 70 days to contain those breaches once they were identified [22].

Malicious software (malware) is one of the most common methods adversaries use to exploit computer networks. Generally, malware-based attacks use manipulated software to intentionally cause damage or access data. A single successful malware attack can result in millions of dollars in damages, with recent annual financial impact measured to be in the hundreds of billions of dollars [3, 12]. Each day, adversaries design new and increasingly complex malware systems, challenging security professionals to deploy robust and effective counter measures.

As a result, malware detectors have become a critical component of a cybersecurity strategy. Traditionally, anti-virus systems used signature-driven systems (i.e., systems that looked for specific software known to be malicious) to detect malware [28]. More recently, dynamic software analysis has become increasingly popular, where evaluated software is run in a secure environment to directly observe whether or not it behaves maliciously [10]. While powerful, anti-virus and dynamic analysis tools have limitations. These methods can be (1) time consuming, when the data volume and velocity are high [25], and (2) operationally intractable, when access to original files may be restricted due to data sensitivity, storage limitations, or the nature of tools themselves. Therefore, organizations may use analytic methods that are based only on file metadata to prioritize deeper analysis or even as the sole method of analysis. While this can increase the efficiency of cybersecurity operations under most conditions, it is also essential when the file itself cannot be examined.

For example, during incident response, cyber protection teams (CPTs) may need to evaluate data on a disconnected, edge network, where connectivity restrictions may prohibit the exfiltration of data into an enterprise analytic environment. Under these conditions, CPTs will bring separate analytic tool kits, with limited compute, to securely assess data using techniques that will not expose the kits themselves to compromise (e.g., without accessing raw executables). Methods must evaluate large volumes of data using inference engines that are computationally-efficient and based on the minimal amount of information possible, mitigating compromise expansion risk and rapidly flagging the most-likely malware culprits.

Additionally, organizations may deploy these methods when:

(1) New external hard drives or thumb drives are connected to a network computer. There may be thousands of new files on these external drives, where comprehensive scans will take time and slow down the drive, potentially interfering with work.
(2) Network operators want to sample files from the network to check for potential malicious content. Downloading all files off of any individual machine may put too much strain on the network, meaning only a subset of files can be selected from any machine. Sampling files randomly from any given machine is unlikely to be effective, as recent work has estimated benign files to outnumber malicious ones at a ratio of 80:1 [18].
(3) Security professionals begin a forensic investigation into destructive malware or corrupted storage media, In many cases, it may not be possible to recover working or complete versions of the files on a system. File metadata is often more likely to remain intact.

Therefore, in this work, we aim to explore whether file names are predictive of malware. Though evaluating maliciousness based on file names may not be definitive, it may help inform investigations and support organizations when complete file analysis is intractable.

For many reasons, the predictive nature of filenames is a reasonable hypothesis. First and foremost, naming software in a descriptive manner is generally considered a “software engineering best practice." As such, we suspect the habitual behavior of software developers may lead to valuable information leakage into file names. Furthermore, for small-scale, targeted malware attacks, file names are likely to be well-designed by adversaries to mask the malware. However, in large-scale malware attacks, where the manual specification of file names is not feasible, adversaries are likely to use algorithmic name generation techniques using structured routines. This may create patterns in malicious file names.

In this work, we demonstrate that a file's name contains predictive information about its maliciousness. While we do not propose this method as a replacement for anti-virus solutions, it can be a valuable tool for security professionals when resources or information are constrained. The related work to our own will be discussed in section 2, followed by the data used in section 3. Our methodology will be discussed in section 4, where we specify the models we evaluate and the design choices for our solution. In section 5 we will review the results of our method, where we show that file names are surprisingly accurate and investigate their utility. To confirm our method has not inadvertently relied on inappropriate information, we perform a cluster and visualization analysis of our model in section 6. Finally, we will conclude in section 7.

## 2 RELATED WORK

Using contextual information alone to predict a file's maliciousness has been a small but relatively successful endeavor. Two primary approaches have been used. Chau et al. [6] introduced file-source context, where the reputation of both files and machines are combined to make a contextual graph. Propagating through this graph can infer benign/malicious likelihoods for new files based on the machines on which they are found. Ye et al. [29] introduced an approach based on file co-occurrence, as new files are often deployed onto a machine in groups, coming from a similar archival file (e.g., a zip, tar, or msi file). Here, archive source information and individual file maliciousness is used to create contextual graphs, which information is once again propagated through to infer maliciousness for new files.

Both graph-based approaches have been improved further, [11, 16, 27] as they are valuable to network defense. Though they enable file evaluation and prioritization with efficiency and scale, they also experience limitations. The information needed is generally only available to anti-virus companies with large repositories of data, making external replication and implementation difficult. Our approach of using file names is the first contextual feature that can be readily replicated by others, as file name information can be obtained publicly from VirusTotal reports [1], which are included with the public EMBER dataset.

Demetrio et al. [9] and Coull and Gardner [8] have both evaluated the efficacy of deep learning methods, namely [23], for malware detection from raw byte code. While they reach diverging conclusions over algorithmic learning, both agree that metadata information, which does not necessarily cause maliciousness, is used to make predictions. This similarly suggests non-causative information that is useful for prediction may be present in file names.

Simultaneously with our work, a pre-print by Kyadige et al. [17] demonstrates that malware detection can be improved through the examination of the file *path* context[1]. In particular, they show that deep neural networks benefit from file path context when predicting maliciousness, even though file path structure is not necessarily related to the malware itself.

Though our work uses similar information as the aforementioned research, there are a number of pertinent differences. First, by using file paths as a foundation, Kyadige et al. rely on information that is not generally accessible (outside of anti-virus companies with large customer sets). Our use of only file *name* (e.g., just "evil.exe") makes our work immediately reproducible with a Virus Total report. The use of file paths may also be limited in some circumstances (e.g., when a file intercepted over a network does not yet have a file path, though it does have a file name), and the subsequent methods may be sensitive to where a file resides (e.g., on an external drive or after a file was copied to a system). Generally, Kyadige et al. focus on the evaluation and design of a full anti-virus system, while our work is motivated by different objectives, as discussed in section 1.

## 3 DATA

We use the EMBER data set from Anderson and Roth [2], an open source data set of over a million portable executable file (PE file) sha256 hashes scanned by VirusTotal in 2017. The data set includes metadata, features derived from the PE files, and a benchmark model trained on those derived features. In addition to the publicly available EMBER data set, we used the reports generated by VirusTotal to extract the file names associated with each PE file, which are found under the `submission_names` field of the reports. Data lacking file names were dropped, and when multiple names were associated with a file, the first was used. Additionally, unlabeled data in the training set was dropped. After initial analysis, we found a number of files that where easily classified in bulk due to malware related names being appended to the original name. To avoid biasing our results with potentially unrealistic true-positive rates, we dropped data with file names containing substrings `vir`, `mal`, and `hack`. In these cases, it is possible that VirusTotal users uploaded files with names modified to include the malware or virus labels. Further investigation did not reveal other large-scale naming conventions that may have been caused by users uploading known malware.

### 3.1 Data Properties

After processing, the training data consisted of 284,999 benign files and 185,615 malicious files, and the test data consisted of 99,204 benign files and 84,086 malicious files. Example file names for each class can be found in Table 1. Below, we examine some characteristics of the file names found in the training data.

Figure 1 is a histogram of file name character lengths in the training data for each class. It can be observed that the number of files of a certain length decreases exponentially as the length of the file name increases. Benign and malicious files both have similar name length distributions.

Figure 2 shows the frequencies of the characters found in the training data for each class. Benign and malicious files have nearly

[1] e.g., `C:\Documents\Files\evil.exe`

identical frequency characteristics. It can be observed that the majority of the file names are composed of a small number of characters.

In the training set, 307 unique characters appear in the malware class that do not appear in the benign class, and only a tiny fraction, 0.11%, of the malware instances contain one of these malware-unique characters. Conversely, 621 characters appear in the benign class that do not appear in the malware class, and 0.10% of the benign instances contain one of these benign-unique characters. 417 characters appear in both classes. We note that only about a tenth of a percent of the training data contains one of these class-unique characters, and most of the class-unique characters are non-ASCII special characters or characters from non-Latin writing systems. The low frequency and the makeup of these class-unique characters suggests that their presence in the data is not abnormal, given that the EMBER data set was built using files submitted to VirusTotal.

A vocabulary size $V_{size}$ is specified at training time to limit the characters processed to the $V_{size}$ most frequent characters. At prediction time, characters that are not part of the vocabulary are ignored.

**Table 1: Example file names for each class.**

| | File Name |
|---|---|
| Benign | d3d9.dll |
| | e5b355a6c7cb37ba88f6a6daa61fce5f |
| | iexplore.exe |
| | wmiutils.dll |
| | winscp |
| Malware | ctvqzym.exe |
| | adobe_epic.dll |
| | img002.exe |
| | servicess |
| | Sennepsfabrikkernes0 |

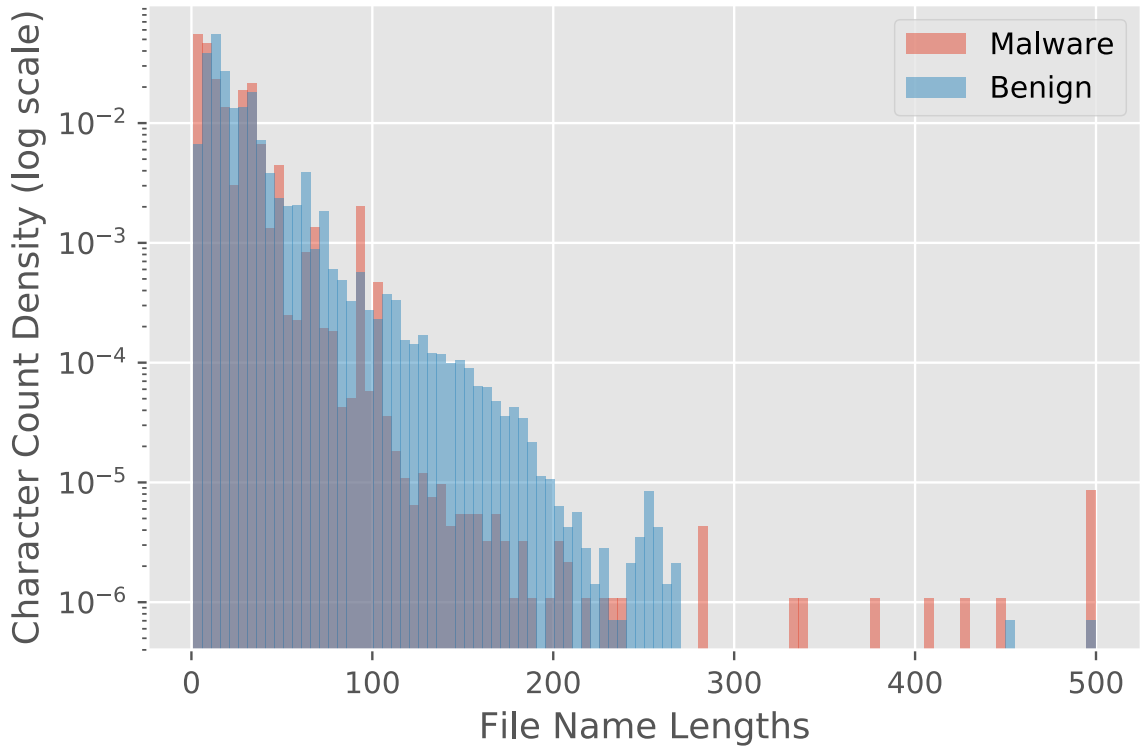

**Figure 1: Histogram of file name character lengths in the training data. The count densities are log scale. The number of file names of a certain length decreases exponentially as the length increases.**

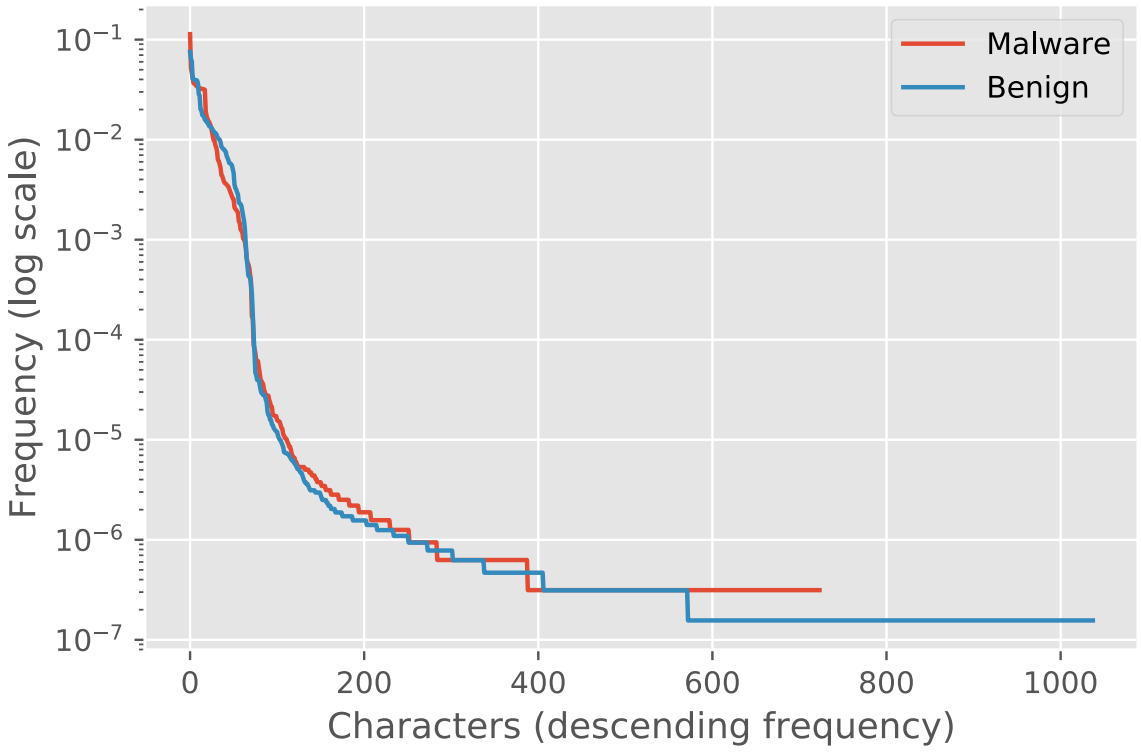

**Figure 2: Frequency of the characters found in the training data. The frequency is log scale.**

# 4 METHODOLOGY

In this work, we evaluate two models that are based solely on file names. First, as a base-line, we consider a Logistic Regression classifier. While it performs well, we see that greater accuracy can be achieved by applying a Convolutional Neural Network (CNN). As file names are neither causative nor directly informative of malicious behavior and can be easily modified for adversarial attacks [4], it would not be wise to use these methods for comprehensive malware defense. Rather, this method would be used to prioritize deeper file evaluation or to act as an indicator of compromise (IOC) when raw files are unavailable. A normal anti-virus product, created for comprehensive defense, would be evaluated based on recall at low false-positive rates (e.g., $10^{-3}$ to $10^{-5}$). As our method has alternate objectives, we use Area Under the ROC Curve (AUC) as our primary performance metric. Generally, this score corresponds to the quality of a ranking of all malicious files with a score higher than all benign files, which is an appropriate criterion for our application [7]. We also consider the F1-score as a generic balance between precision and recall, as we may not necessarily prefer one over the other (in contrast to a normal anti-virus product).

## 4.1 Logistic Regression

As baselines, we use a binary 3-gram character embedding in combination with a L2-regularized logistic regression and a L1-regularized logistic regression. Both methods are characterized by the objective function in Equation 1, where $R(w) = \|w\|_1$ produces L1 regularization, and $R(w) = \frac{1}{2}\|w\|_2^2$ results in L2 regularization.

$$f(w) = R(w) + C \sum_{i=1}^{N} \log\left(1 + \exp\left(-y \cdot w^{\top} x_i\right)\right) \quad (1)$$

The 3-gram character embedding results in a 163177-dimensional feature space. The sparsity of the vectorized training data is 99.99%, where sparsity is defined as the number of zero-valued elements divided by the total number of elements in the data matrix. The sparsity of the data urges L1-regularization, which tends to improve predictive performance on extremely sparse data [21]. We also evaluate L2-regularization for completeness.

The logistic regression regularization strengths were optimized using cross-validation, resulting in a L2-regularization strength of 0.36 and a L1-regularization strength of 2.78.

## 4.2 Character-level Convolutional Network

To investigate the efficacy of a higher capacity model, we adapt the character-level convolutional neural network (CharCNN) architecture from Zhang and LeCun [30]. CharCNNs operate on text data directly at the character level, making them particularly powerful. They do not require knowledge of the language's syntactic or semantic structure nor the specification of a representation (such as words or n-grams). In particular, we use the small architecture described in Zhang and LeCun [30], and we modify the architecture by replacing the one-hot encoding of the characters with a learned character embedding layer of identical dimensionality. Table 2 summarizes the neural network architecture we use.

We use cross entropy loss as the cost function for training over a total of 10 epochs, using the Adam optimizer with a learning rate of 0.001 [15]. The character vocabulary size was set to 300 (i.e., character tokens not among the 300 most frequently occurring characters were ignored), and the text window size was set to 100 (file names were truncated to length 100). Both are reasonable per Figure 2 and Figure 1, as these settings capture most of data.

**Table 2: CharCNN architecture. The network consists of three sections: (1) a character embedding section, which replaces the one-hot encoding from Zhang and LeCun [30], (2) a convolution section, and, (3) a fully connected section. Rectified linear units (ReLU) are used as non-linearities after each layer, with the exception of the last layer, and 0.5 probability dropout is used after the first two fully connected layers.**

| | Layer | Feature / Output | Kernel | Pool |
|---|---|---|---|---|
| Embedding | 1 | 300 | | |
| Convolutional | 2 | 256 | 7 | 3 |
| | 3 | 256 | 7 | 3 |
| | 4 | 256 | 3 | |
| | 5 | 256 | 3 | |
| | 6 | 256 | 3 | |
| | 7 | 256 | 3 | 3 |
| Fully Connected | 8 | 1024 | | |
| | 9 | 1024 | | |
| | 10 | 2 | | |

*4.2.1 Combining File Names with Ember Features.* We also evaluate fusing file names with the original EMBER features. For this experiment, we use a two hidden layer multi-layer perceptron, where the number of nodes in each hidden layer is equal to the dimensionality of the input data (i.e., 3375 nodes). ReLU [20] is used as the non-linearity, and each hidden layer is followed by batch normalization[13]. As input data, we concatenate the original EMBER features with the file name embeddings learned by the CharCNN (i.e., the output of layer 9 from Table 2).

# 5 RESULTS AND DISCUSSION

In this section, we examine and discuss our results. For comparison, metrics for the LightGBM model released with the EMBER data set and pre-trained on the original features are reported in Table 3 [2, 14]. We do not expect malware detectors based solely on file names to perform as well as the LightGBM model trained on the original features in the data set, as those features contain information from file contents extracted with domain knowledge of the PE file format. This makes the EMBER features more information rich than the file name alone. We emphasize that the goal of this work is to demonstrate a method that could be used in parsimonious situations where file contents may not be available to inspect.

**Table 3: Metrics for the LightGBM model released with the EMBER dataset and pre-trained on the original EMBER features.**

| | Precision | Recall | F1-score |
|---|---|---|---|
| Benign | 0.99 | 0.99 | 0.99 |
| Malware | 0.99 | 0.98 | 0.98 |
| Average | 0.99 | 0.99 | 0.99 |

## 5.1 Logistic Regression Results

Table 4 and Table 5 summarize the logistic regression results. As expected, L1-regularized logistic regression performs slightly better than its L2-regularized counterpart. A comparison with Table 3 shows that, while the performance of logistic regression on file name character 3-grams is not as strong as the LightGBM model trained on a wider set of features, performance remains very competitive. Specifically, the L1-regularized logistic regression on file names alone maintained 92% of the accuracy achieved by the LightGBM model, as measured by average F1-score. This alone validates the notion that file names have valuable information for prediction. Figure 3 shows the receiver operating characteristic for logistic regression, where the AUC is 0.97 for both regularization schemes.

**Table 4: Metrics for the L1-regularized logistic regression on file name character 3-grams.**

| | Precision | Recall | F1-score |
|---|---|---|---|
| Benign | 0.89 | 0.95 | 0.92 |
| Malware | 0.94 | 0.86 | 0.90 |
| Average | 0.91 | 0.90 | 0.91 |

**Table 5: Metrics for the L2-regularized logistic regression on file name character 3-grams.**

| | Precision | Recall | F1-score |
|---|---|---|---|
| Benign | 0.88 | 0.95 | 0.91 |
| Malware | 0.94 | 0.84 | 0.89 |
| Average | 0.91 | 0.90 | 0.90 |

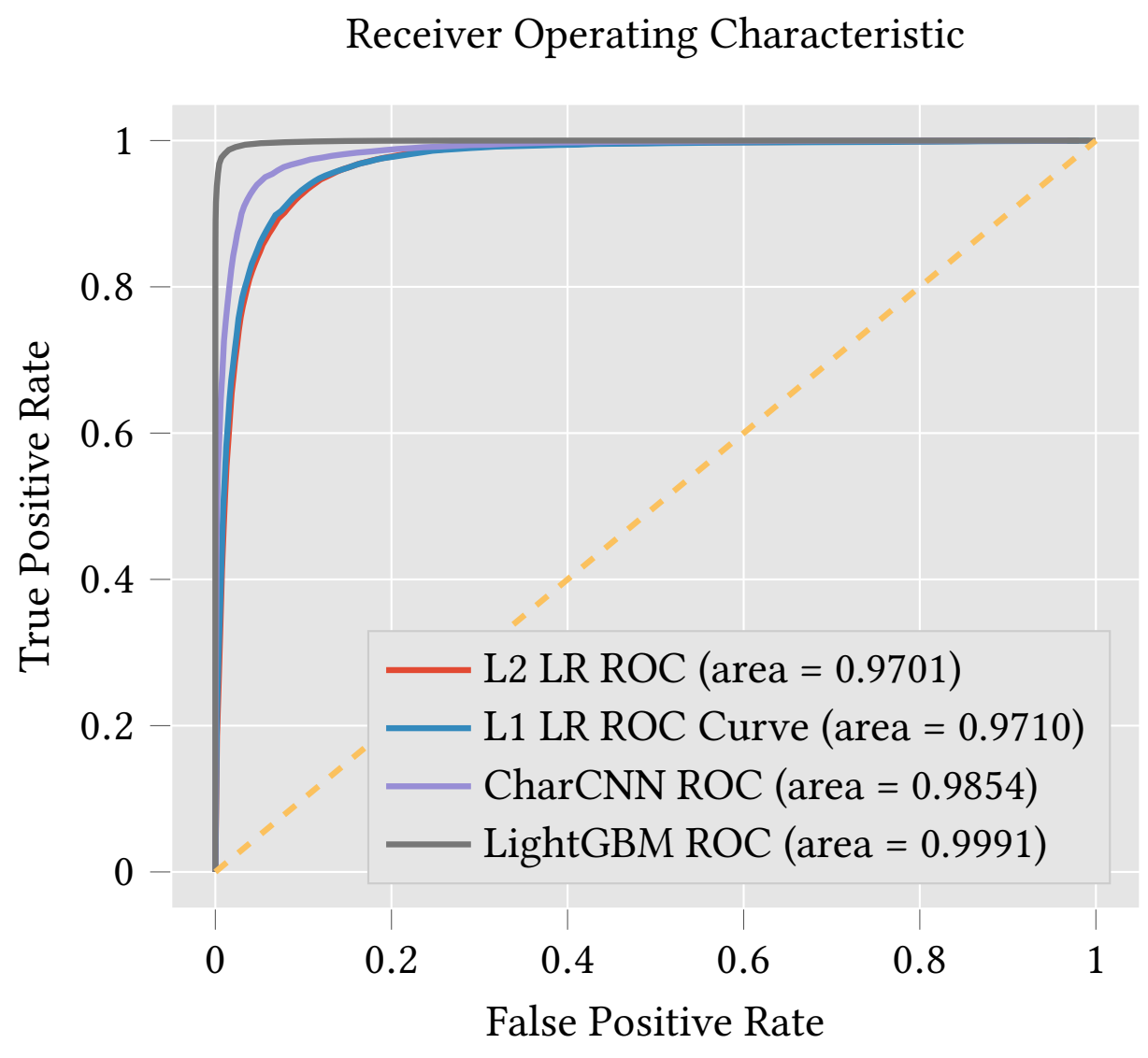

**Figure 3: Receiver operating characteristics.**

The logistic regression weights can also be examined to uncover the character 3-grams most indicative of both benign and malicious files. Table 6 shows the most indicative features for both classes. We note no obvious indicators of maliciousness in the features shown to be the most predictive of malware. If we had not removed strings containing `vir`, `mal`, and `hack`, they would have dominated this list of features. Accordingly, we have reasonable confidence in our removal of severe information leakage.

**Table 6: Character 3-grams most indicative of benign files and most indicative of malware, based on learned L1-regularized logistic regression weights.**

| | Features |
|---|---|
| Benign | 'en/', 'ub3', '.so', 'WAX', ' i0', 'pyd', '.fi' 'xes', 'sdm', '.ef', 'sp7', '.mu', '.oc', '1b.', 'jpl' |
| Malware | 'Ac.', 'tdL', 'kns', 'xtC', 'cpS', 'Pe ', 'MsU', 'pym', ' 1.', 'nd3', 'UBU', ' .e', 'bi ', 'HMD' |

## 5.2 Character-level Convolutional Network Results

Table 7 summarizes the results of the CharCNN based solely on file names. A comparison with Table 3 shows that, while the performance of the CharCNN is not as strong as that of the LightGBM model trained on a wider set of features, the CharCNN achieves 96% of the accuracy achieved by the LightGBM, as measured by average F1-score. Figure 3 shows the receiver operating characteristic for the CharCNN model where the AUC is 0.99. A comparison with Table 4 and Table 5 indicates that the CharCNN improves upon logistic regression across all metrics.

The results align with our expectations that EMBER's original domain knowledge features would provide the best performance, though the CharCNN's accuracy is more than sufficient to warrant the method's application to the use-cases we are interested in. A relatively high AUC of 0.985 implies that, in a rate-limited situation, the method will ensure security professionals will have the opportunity to process most all malware before most benign files. An F1-score of 0.95 is similarly encouraging, and we argue that this method would be a strong IOC for investigative use, particularly in data-constrained situations.

**Table 7: Character-level convolutional neural network metrics.**

| | Precision | Recall | F1-score |
|---|---|---|---|
| Benign | 0.94 | 0.96 | 0.95 |
| Malware | 0.95 | 0.93 | 0.94 |
| Average | 0.95 | 0.94 | 0.95 |

At prediction time, the CharCNN takes on average 1.69 milliseconds of processor time to label a batch of 64 using a single NVIDIA Tesla V100 GPU.

## 5.3 Combining File Names with Ember Features

Table 8 summarizes the results from combining the CharCNN embedded file names with the original EMBER features using a MLP. As shown below, the method maintains an average F1-score of 0.97, with an absolute improvement of 0.02 compared to the original CharCNN method discussed previously (see Table 7). At first impression, fusing these two feature vectors improves the method and approaches the performance of LightGBM on the original features (see Table 3).

**Table 8: Metrics for MLP trained on the original EMBER features, combined with the CharCNN embedded file names as additional features.**

| | Precision | Recall | F1-score |
|---|---|---|---|
| Benign | 0.96 | 0.98 | 0.97 |
| Malware | 0.98 | 0.95 | 0.96 |
| Average | 0.97 | 0.97 | 0.97 |

For completeness, we also train the same neural network architecture on just the original EMBER features. This method performs similarly to the LightGBM, with an F1-score of 0.99, as shown in Table 9. Therefore, including the CharCNN embeddings actually reduced the effectiveness and accuracy of our model!

**Table 9: Metrics for MLP trained on only the original EMBER features.**

| | Precision | Recall | F1-score |
|---|---|---|---|
| Benign | 0.99 | 0.99 | 0.99 |
| Malware | 0.98 | 0.99 | 0.99 |
| Average | 0.99 | 0.99 | 0.99 |

This result suggests that the file names provide no additional predictive value compared to information that can be parsed out of the binaries and used as features. Whatever information that is leaked into the name about the file's nature (benign/malicious) is simply a redundant and *noisy* version of information available in the binary. As this version of the information is of increased noise, it is less reliable when used by the neural network, and thus reduces overall performance.

This contrasts with the results reported in Kyadige et al. [17], who found file paths combined with file features improved accuracy. We have two hypothesis that could explain this result. First, the complete file path may contain more information than just file names and is thus more reliable. We do not suspect this fully explains the difference, as the feature vector for file path context is only 100 dimensions and does not contain the full path in its original form.

Our second hypothesis is that the file-based features used by [17] simply do not contain the same level of information as the extracted features used in the EMBER corpus. They use the same feature approach as the seminal work by Saxe and Berlin [26], which uses a histogram of entropy values, a histogram of string lengths, a histogram of entropy standard deviations, and 256 dimensional bin to hash values extractable from the PE-header. This last set of 256 features will have collisions, as it corresponds to feature-hashing into a small dimensional space. The 2,351 EMBER features have these histograms, in addition to keeping the domain knowledge explicitly represented with their own features, avoiding unnecessary disambiguation. It seems more likely that their improvement is subsequently a function of re-introducing a (noisy) version of information that was lost due to unnecessary feature hashing.

## 5.4 Interpretation of Results

Our results show that file names can contain predictive information about its nature (benign/malicious). While there is not necessarily an inherent relationship between malware and the name of the file it lives in, information about malware status likely leaks into the file name as a result of how the malware programmer chooses the file name. This suggests that screening files using file names may be effective in certain situations and would be valuable when it is not computationally feasible to run a full malware analysis on all of the data. Additionally, the use of file names to flag potentially malicious software is useful in situations where access to the underlying file is restricted. For example, an analyst may be in an environment with highly sensitive data or may be using tools that do not include access to or do not store original files.

We note two main limitations of our analysis. First, the EMBER data set is considered a relatively easy data set where, as shown in Table 3, 99% accuracy can be achieved using out of the box features and tree-based gradient boosting algorithms.

Second, a malware detector of this type would be very easy to evade from an adversarial machine learning perspective, as the malicious attacker has considerably less constraints when modifying file names than when modifying the actual file contents. In particular, the attacker does not need to ensure that, post-adversarial perturbation, the file still behaves as intended when modifying just the file name. This may not be as debilitating to our approach as it first sounds - the effort required to evade our approach may not be worthwhile to attackers. As our approach would not cover the majority of cases nor replace anti-virus detection on endpoints, it is most useful specialized and niche cases. Second, many applications of our method are for the prioritization of files by a more complete anti-virus system later. As such, it may be possible to deploy a Multi-Armed Bandit approach to adapt to multiple different sampling strategies (e.g., our file name approach and random sampling), and use the more complete anti-virus results to determine which sampling strategy is performing best. This depends on the assumption of down-stream systems being more robust, which we argue is a reasonable assumption, leaving more complete evaluation to future work.

# 6 QUALITATIVE EVALUATION

It may seem unusual or surprising to some that file names are so predictive of malicious behavior. Thus, it is desirable to investigate our model beyond the cursory evaluation of test set accuracy. Is it possible that data biases result in simple file names being repeated across training and testing data? Are there real systemic flaws or patterns in file naming conventions that are exploitable, or does one of the classes exhibit a degenerate naming convention that makes it trivially separable from the other class? We investigate these questions using unsupervised visualization and clustering techniques, and conclude that there is no obvious systemic issue leading to inflated accuracy scores. It appears that there are indeed patterns in file naming behaviors that are leaked by software design and development, which we can leverage in the aforementioned use cases.

For this qualitative exploration, we note that the CharCNN's penultimate fully connected layer's output can be viewed as an embedding of the file names (where benign files and malware are hopefully well separated). After training, we compute this embedding for all of the training data and compress the representation from 1,024 dimensions to 2 dimensions using a Uniform Manifold Approximation and Projection (UMAP). A popular tool for visualization of high dimensional data and for general non-linear dimension reduction, UMAP is a manifold learning technique constructed from a theoretical framework based in Riemannian geometry and algebraic topology [19].

Figure 4 visualizes the UMAP of CharCNN learned embeddings on the training data. We note a couple of dense and well-separated malicious clusters and a general trend toward benign and malicious files having homogeneous neighbors. However, the UMAP embedding should not be taken as absolute truth as to the nature of the data.

To further examine the properties of the data, we also cluster the entire training set using the CharCNN's penultimate fully connected layer's output as a representation for the data. HDBSCAN was used as the clustering algorithm. HDBSCAN is a density-based, hierarchical clustering algorithm which improves upon DBSCAN [5]. We use HDBSCAN because (1) it maintains high cluster stability, (2) it does not require the number of clusters to be specified beforehand, and (3) it handles outlier detection naturally.

HDBSCAN found 9,051 clusters in the training data, with 32% of the data labeled as noise with respect to the defined clusters. The true label breakdown of the data points tagged as noise was 111,378 benign and 41,112 malicious. By creating UMAP embeddings of individual clusters, we can gain further insight into the behavior of our models. For example, we plot clusters of three different types in Figure 5, one containing all benign files, one containing all malicious files, and one with a mix.

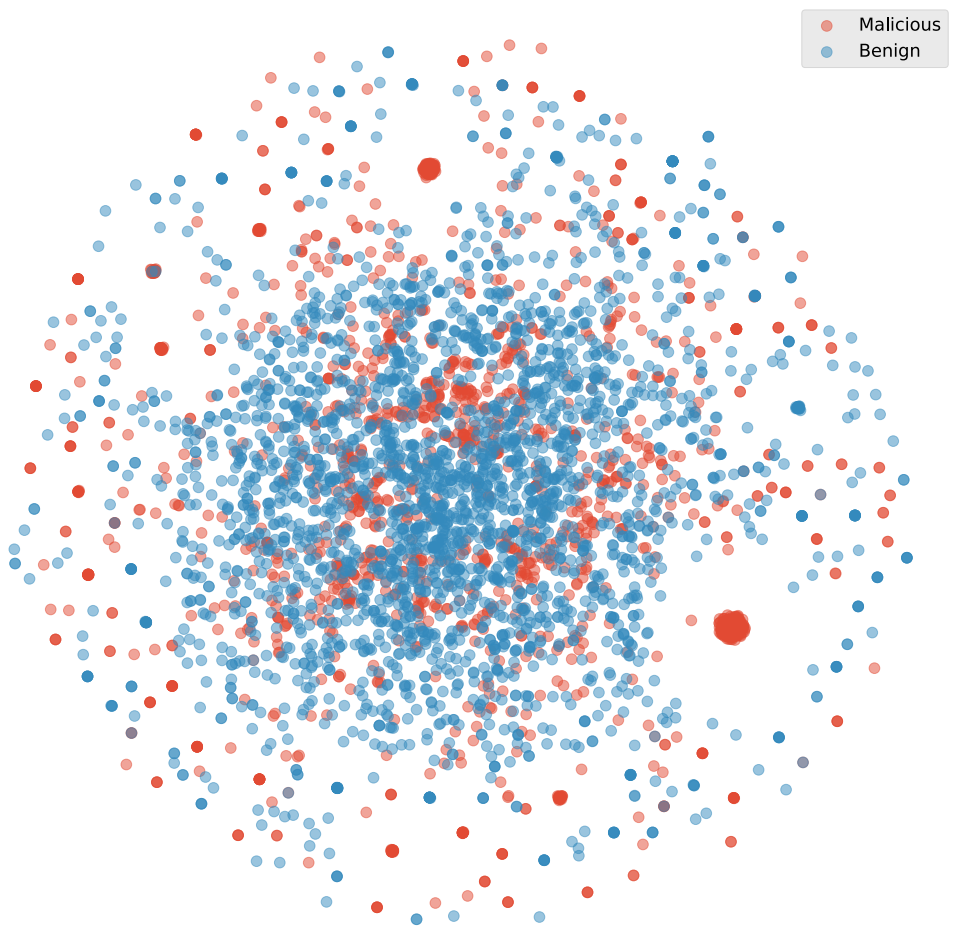

**Figure 4: Visualization of UMAP of CharCNN learned embeddings on the training data. The data was compressed using UMAP from 1,024 dimensions to 2 dimensions. The UMAP embedding was computed using the entire training dataset, and every 50th data point was plotted in this figure.**

In Figure 5a we see CharCNN has isolated a naming scheme that is apparently used by one type of malware. This particular type of malware names itself to pose as a backup program, preceded with a random character sequence. The behavior is simple to understand and follow. In Figure 5b we see a cluster with mixed membership that exhibits a wider array of behaviors. Here, a number of file names have varying degrees of human legibility, transitioning from legible names, to long names that are still legible, to illegibly long names of random characters. Last, we have a cluster of only benign files in Figure 5c. While hard to read, this cluster contains file names that are inordinately long and random. The majority of file names either end in `.bin` or in a pattern of dashes with a temp file name, such as `2-260890ee-19ff-4deb-924f-e1adcd4f74ff.dtm00008.temp`. Manual investigation of a number of the clusters in the corpus reveal a diversity of patterns and behaviors.

## 6.1 Quantifying Cluster Properties and Behavior

It is not possible to manually inspect all clusters, and we would like to evaluate their quality in a more quantifiable and empirical manner. In considering the quality of our clustering, we do not expect (or even desire) that only two clusters emerge, one for malicious and one for benign. In fact, this would be a degenerate situation,

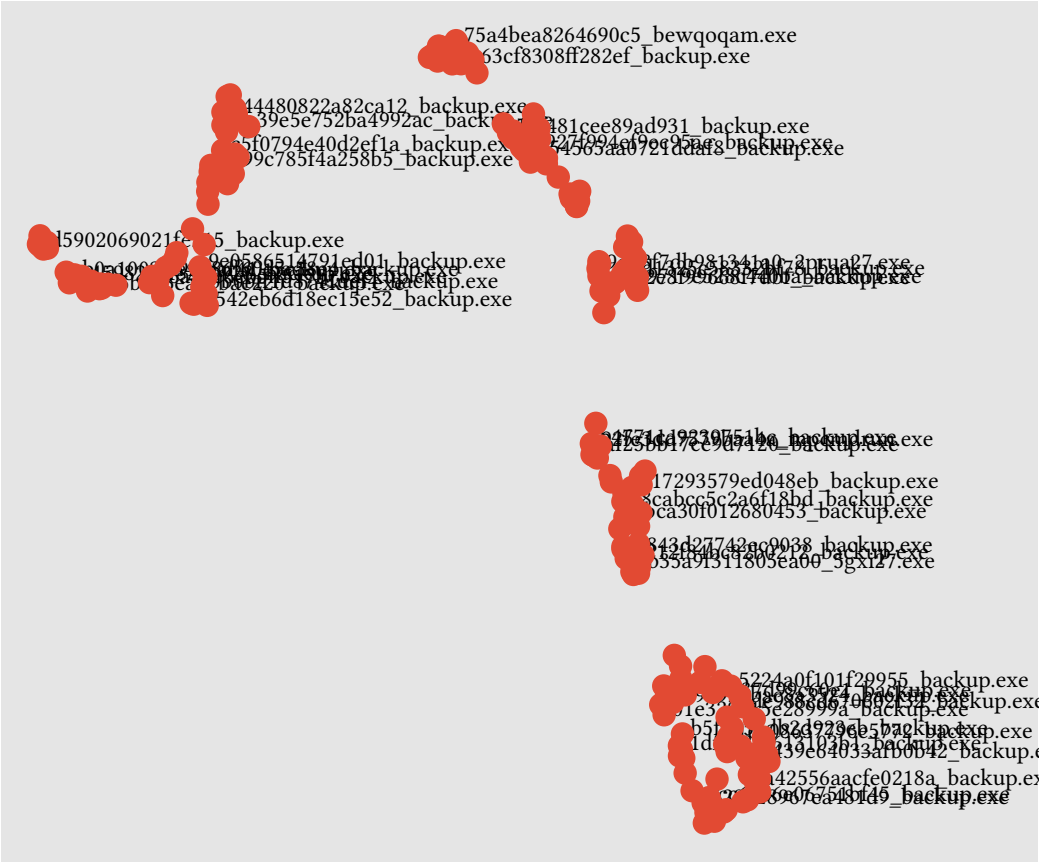

**(a) Malicious Cluster**

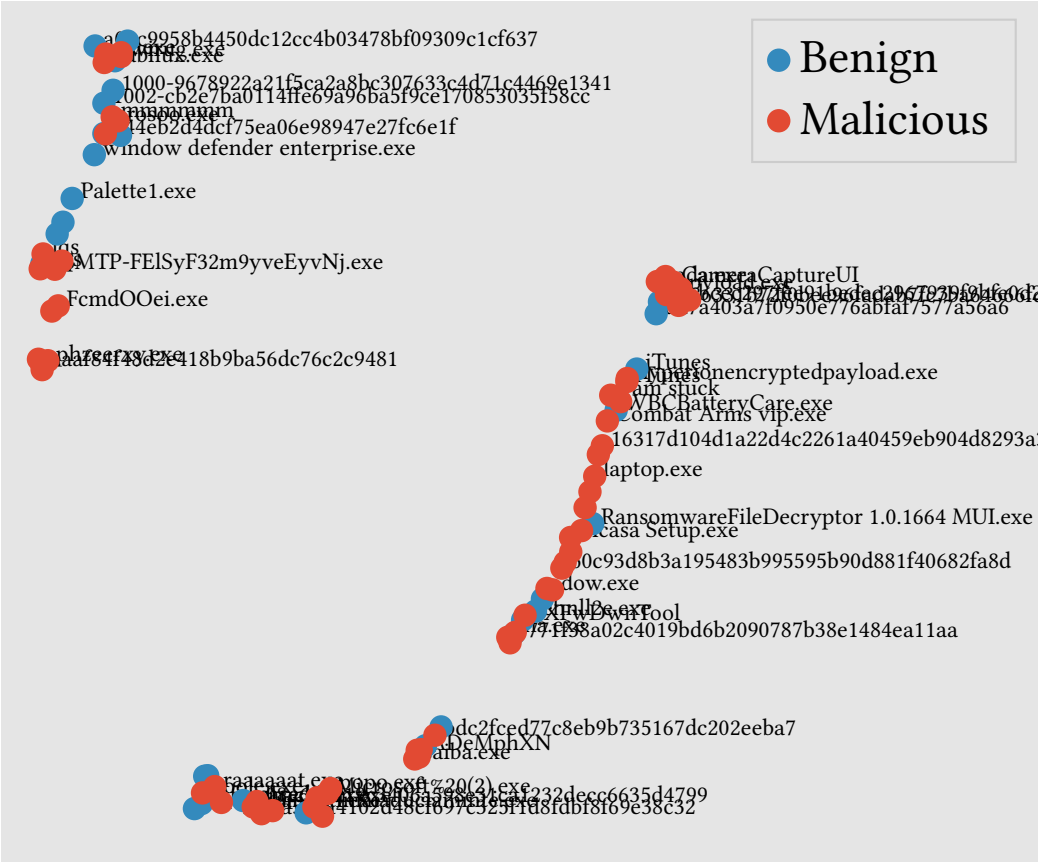

**(b) Mixed Membership Cluster**

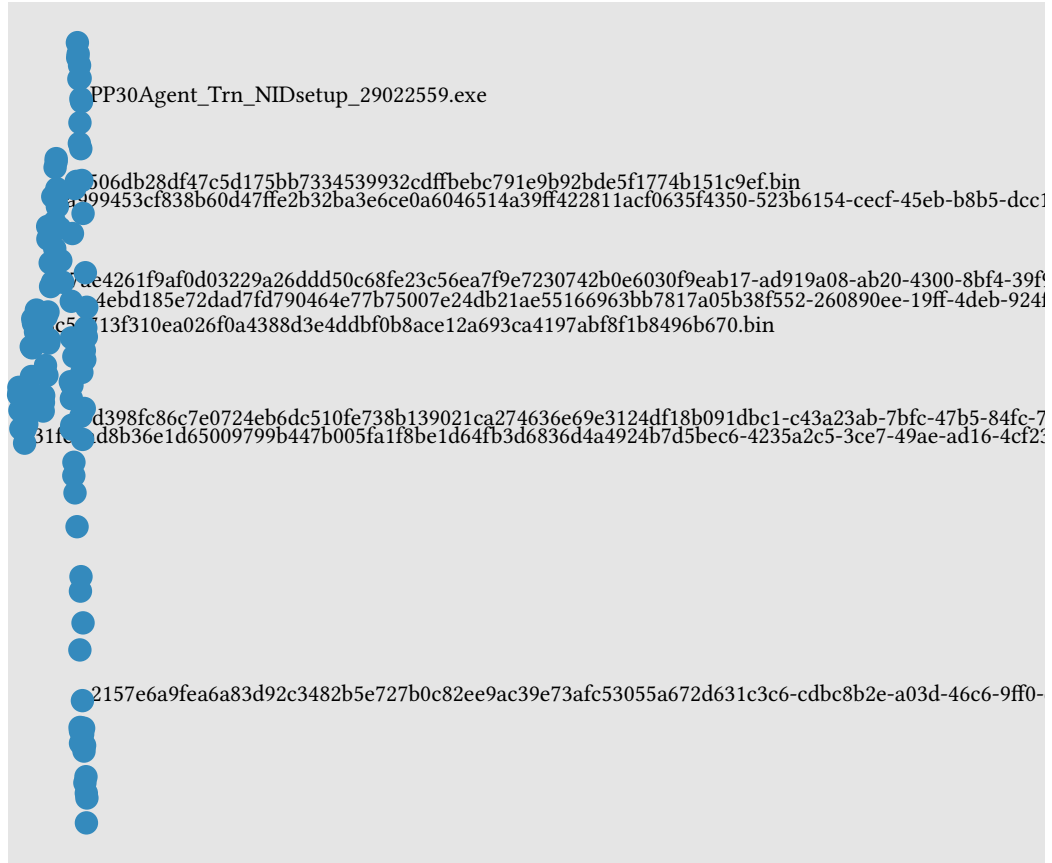

**(c) Benign Cluster**

**Figure 5: UMAP embeddings of different clusters found with HDBSCAN, all plots share the same legend. File names have been placed next to data points to illustrate embedding content (file names selected randomly, with minor removals to improve legibility).**

indicating that the file names in our data are globally homogeneous. This does not align with our expectation that many different types of file names will occur in practice. Therefore, we are concerned primarily with the metric of *Homogeneity* as proposed by Rosenberg and Hirschberg [24], where Homogeneity is measured as the conditional entropy of the class distribution, given the proposed clustering (i.e., $H(C|K)$, where $C$ is the class labels and $K$ the clustering). Homogeneity is maximized ($H(C|K) = 1$) when all clusters contain only one class and minimized when classes are uniformly distributed within each cluster (($H(C|K) = 0$).

The clustering, where outliers are not considered a cluster, achieves a high homogeneity score of 0.916.

This indicates that most of our data is well-grouped and that there are numerous different observable patterns in file naming, which aligns with our expectations.

Figure 6 shows cluster size as a function of cluster maliciousness (i.e., the fraction of the cluster members labeled as malware). We note that the largest clusters, and most of the clusters overall, tend to be more homogeneous (mostly benign or mostly malicious). We also note that the grid-like pattern near the bottom of the y-axis is caused by a finite number of small cluster sizes, where only a discrete number of possible fractions can be achieved. The distribution becomes less rigid as cluster size increases.

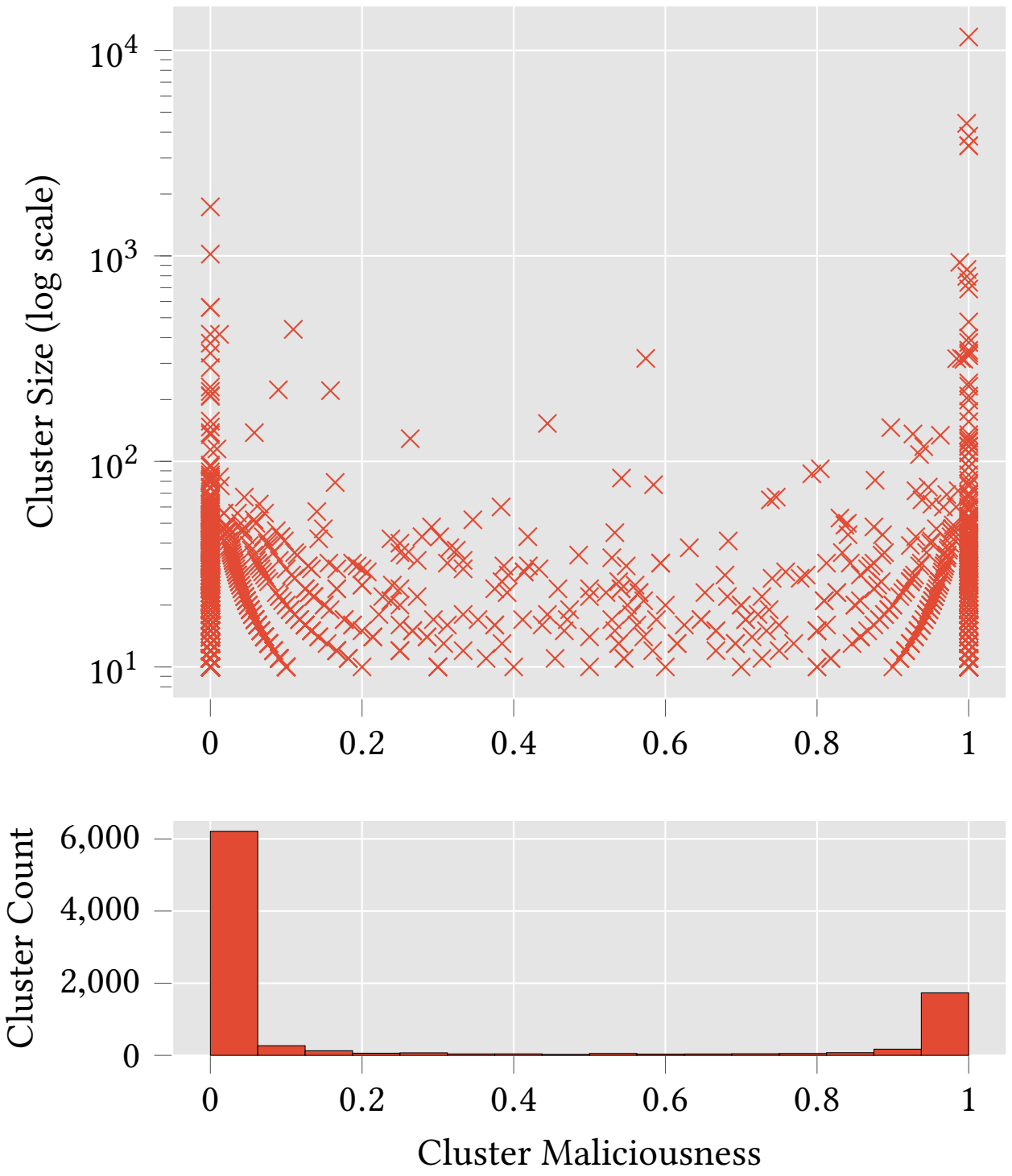

**Figure 6: The top figure shows a scatter plot of cluster maliciousness (fraction of the cluster members labeled as malware) and size. The size is log scale, and the clustering was found using HDBSCAN on the CharCNN's learned representation of the training data. The bottom figure shows a histogram of cluster maliciousness.**

The homogeneous results captured by Figure 6 show many *perfectly* homogeneous clusters. This urges an analysis of cluster composition to ensure the results are valid. In particular, if clusters correspond to a repeated file name within each cluster, there would be concerns around data quality. Since file names come from user submissions to Virus Total, this is a valid concern.

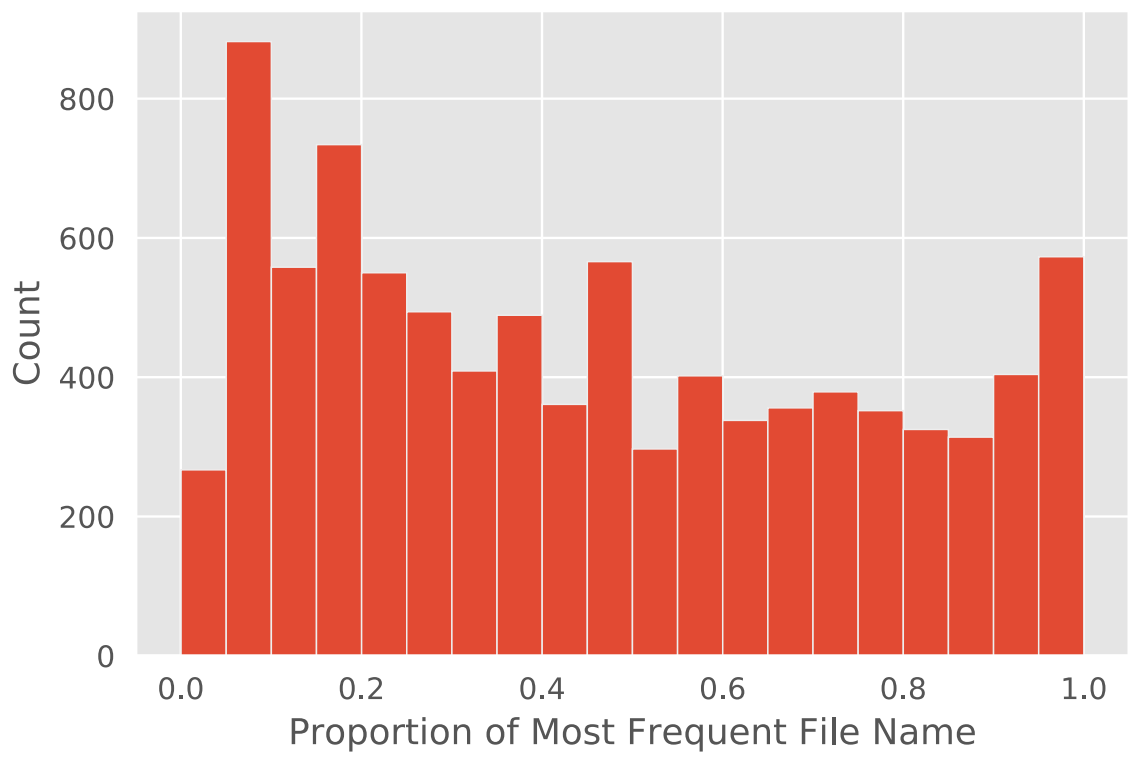

**Figure 7: For each cluster, we computed the proportion of cluster members that correspond to the most frequently occurring file name in that cluster. This figure shows a histogram of the proportions.**

For each cluster, we computed the proportion of cluster members that correspond to the most frequently occurring file name in that cluster. Figure 7 shows a histogram of the proportions. It can be observed that the distribution of proportions is somewhat uniform, suggesting that while there are clusters that correspond mostly to a single file name, there are also more diverse clusters as well. Overall, this indicates that the majority of perfect clusters are not caused by simple, exact duplication of file names. We note that our expectation is that some exact duplication *should* exist. Multiple versions of the same program should have the same name, which would naturally create duplication. A malware author may also use file names of common applications (e.g., Safari, Chrome, Internet Explore), in order to trick victims into running malicious binaries. Thus, duplication in and of itself is not a concern, given the amount of duplication is reasonable and would not prevent generalization.

We also investigate if a relationship exists between cluster duplication proportion and both cluster size and cluster maliciousness, as shown in Figure 8 in Figure 9 respectively. Here, cluster maliciousness is defined as the fraction of the cluster members labeled as malware. Figure 8 and Figure 9 do not indicate any particular relationship between the proportion of cluster members that correspond to the most frequently occurring file name and cluster size/maliciousness.

The presence of clusters with high proportions of members corresponding to the most frequently occurring file name urges the evaluation of the effectiveness of a naive classifier. Here, a naive classifier, during inference, would simply look for an instance of that file name in the training set. If an instance of the file name exists in the training set, the corresponding label is used as the prediction. If an instance of the file name is not found in the training set, the naive classifier returns the label "unseen."

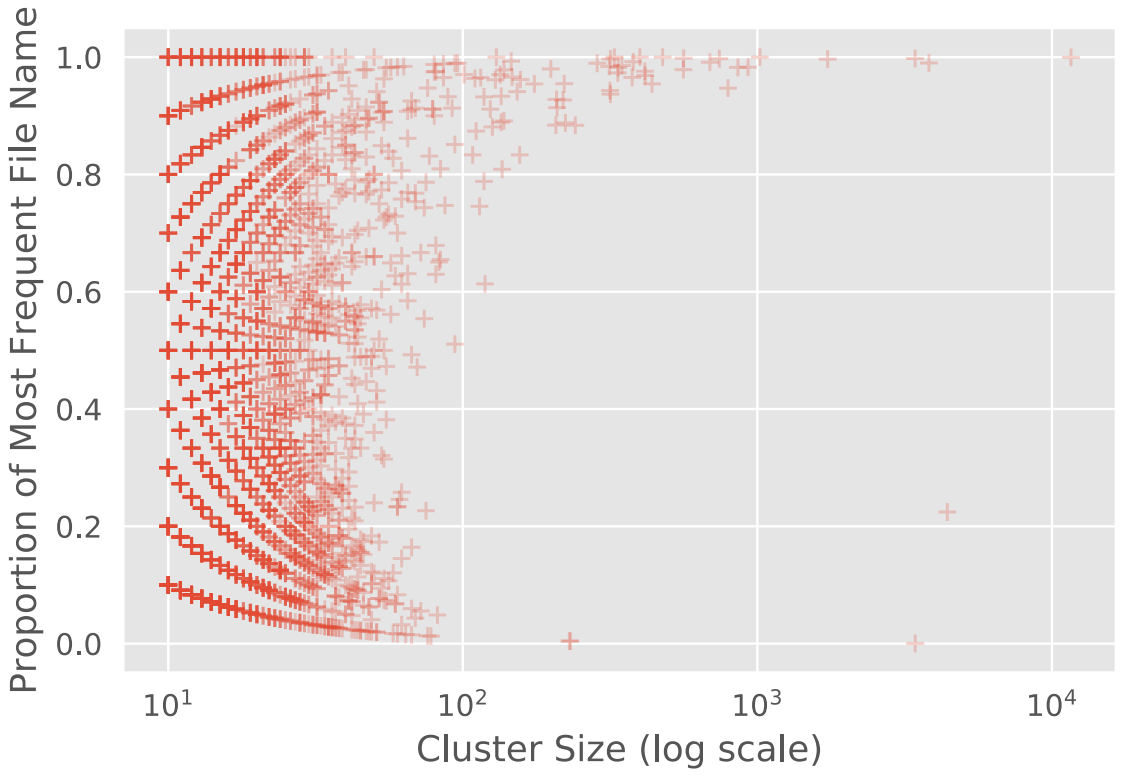

**Figure 8: For each cluster, we computed the proportion of cluster members that correspond to the most frequently occurring file name in that cluster. This figure examines the proportion as a function of cluster size.**

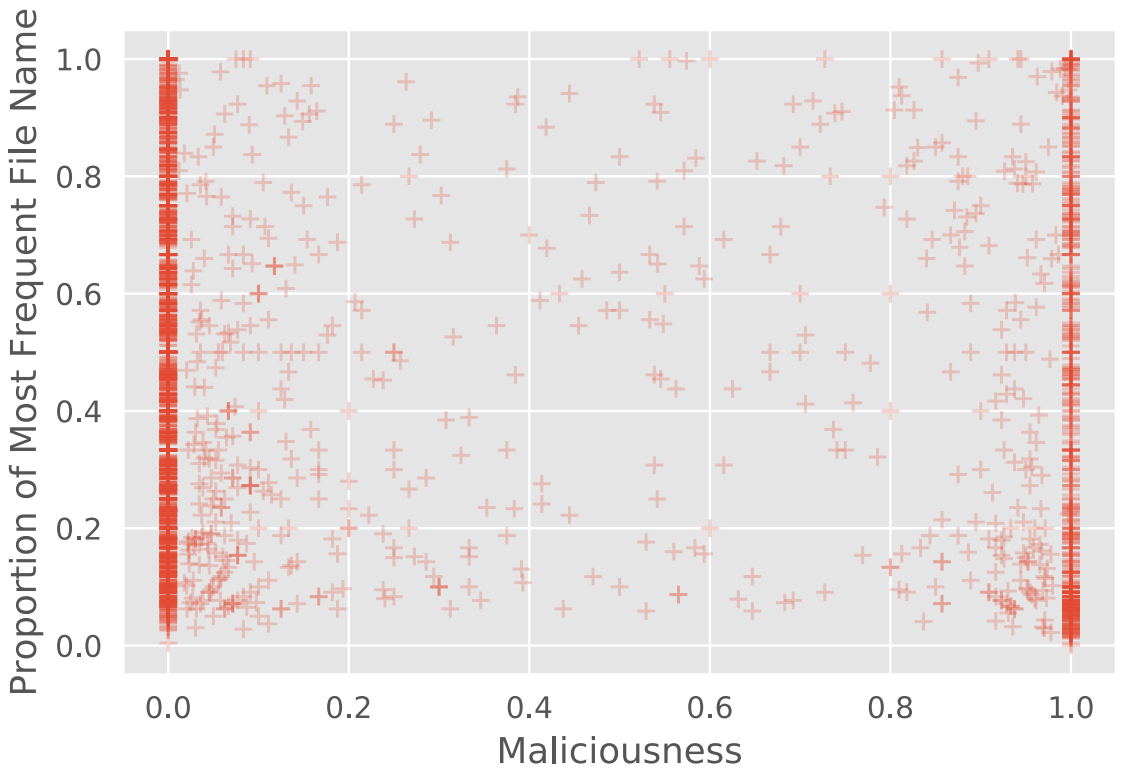

**Figure 9: For each cluster, we computed the proportion of cluster members that correspond to the most frequently occurring file name in that cluster. This figure examines the proportion as a function of cluster maliciousness.**

Figure 10 captures the behaviour of the naive classifier in the form of a confusion matrix, and Table 10 summarizes the classification metrics. We note the high precision but low recall of this naive look up approach. The high precision indicates that if a file name was already seen in the training data, a look up can be quite effective. However, memorizing the training data alone is not sufficient, as the majority of the file names in the test data were not present in the training data. This drives the low recall metrics. Furthermore, this suggests that the high predictive performance achieved by the CharCNN and logistic regression is not due to the memorization of the training data alone.

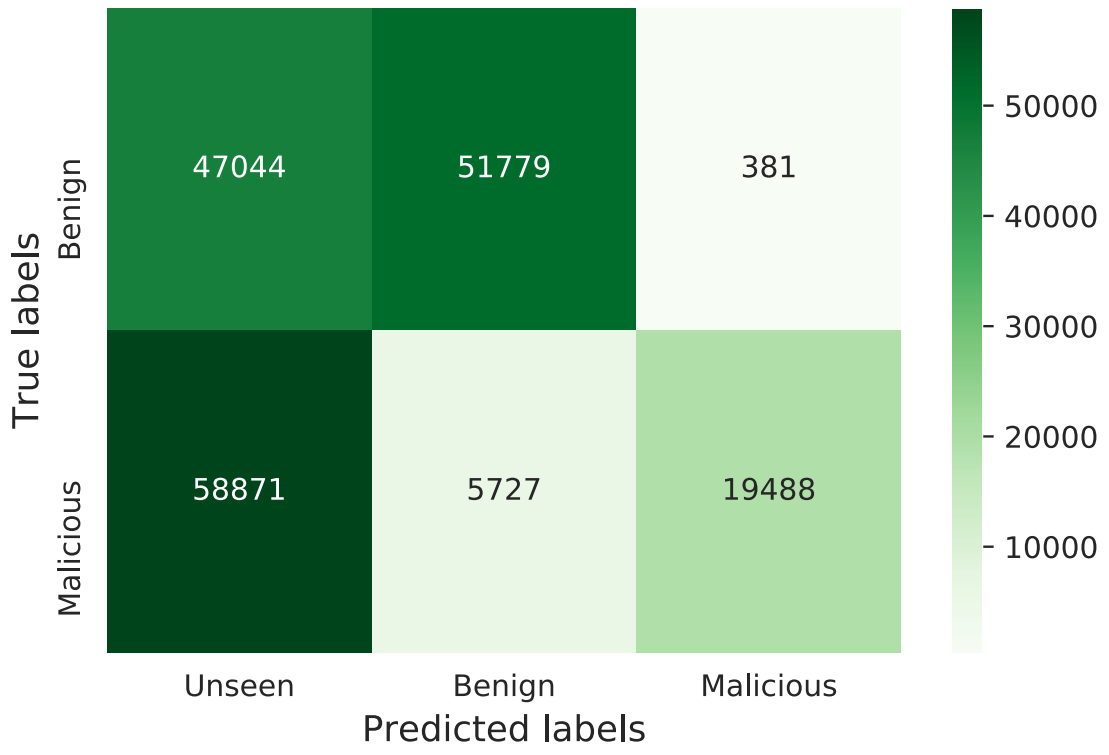

**Figure 10: Confusion matrix evaluating the effectiveness of a naive classifier that, at test time, simply looks for an instance of the test file name in the training set. If an instance of the file name exists in the training set, the corresponding training label is used as the prediction. If an instance of the file name is not found in the training set, the naive classifier returns the label "unseen."**

**Table 10: Metrics for a naive classifier that, at test time, simply looks for an instance of the test file name in the training set, using the corresponding label as a prediction.**

| | Precision | Recall | F1-score |
|---|---|---|---|
| Benign | 0.90 | 0.52 | 0.66 |
| Malware | 0.98 | 0.23 | 0.37 |
| Average | 0.94 | 0.38 | 0.52 |

## 7 CONCLUSION

In this work, we examined the predictive value of file names for malware detection. We established that file names can contain information predictive of malware, and we demonstrated the effectiveness of a character-level convolutional deep learning architecture for the task. We further validated this efficacy through qualitative study of our data and learning methods.